\documentclass[12pt]{iopart}
\usepackage{iopams}

\newtheorem{theorem}[equation]{Theorem}
\newtheorem{remark}[equation]{Remark}

\newcommand{\R}{\mathbb R}

\newcommand{\C}{\mathbb C}

\newcommand{\fii}{\varphi}
\newcommand{\lin}{{\rm lin}}
\newcommand{\<}{\langle}
\renewcommand{\>}{\rangle}
\newcommand{\ket}[1]{\left|#1\right\rangle}

\newcommand{\kb}[2]{\left|#1\right\rangle\left\langle#2\right|}
\newcommand{\rank}{\mathop{\rm rank}}

\begin{document}

\title{Extreme phase and rotated quadrature measurements}

\author{Juha-Pekka Pellonp\"a\"a}

\address{Turku Centre for Quantum Physics\\Department of Physics and Astronomy\\ University of Turku\\ FI-20014 Turku, Finland}

\ead{juhpello@utu.fi}

\begin{abstract}
We determine the extreme points of the convex set of covariant phase observables. Such extremals describe the best phase parameter measurements of laser light --- the best in the sense that they are free from classical randomness due to fluctuations in the measuring procedure. We also characterize extreme fuzzy rotated quadratures.
\end{abstract}


\submitto{Physica Scripta}

\section{Introduction}
Covariant phase observables constitute a simple and elegant solution to the quantum phase problem of a single-mode optical field (see, \cite{Pe} and references therein). They describe coherent state phase (parameter) measurements which can be realized, for example, by using quantum optical homodyne or heterodyne detection. 
Since there exist infinite number of covariant phase observables, it is of great interest to classify the most precise and informative ones.

The set of covariant phase observables is convex. This means that, given two phase observables, one can form a random mixture of them. This mixture describe a new phase measurement. One the other hand, if a covariant phase observable $E$ can be represented as a nontrivial convex combination of two phase observables, one can equally measure these two phase observables and then mix their statistics to get the statistics of $E$.

The aim of this study, is to find such phase observables, so-called {\it pure} or {\it extreme} observables, which do not allow (nontrivial) convex decompositions. 
Pure phase observables then represent the best phase measurements in the sense that they are free from any classical randomness due to fluctuations in the measuring procedure (see, \cite{Ho82}).

Similarly, as in the case of phase observables,  we determine the extreme points of the convex set of fuzzy rotated quadratures. The rotated quadratures are important in quantum optics, since they can be measured by balanced homodyne detection.

The structure of this article is the following: in section 2 we define coherent state phase measurements (of laser light) and the associated phase observables. We also consider the structure of such observables. The canonical phase observable is introduced in section 2.1. A necessary and sufficient condition for extremality of a phase observable is given in section 3.
In section 4, we define fuzzy rotated quadratures and find extremal quadratures.

\section{Phase measurements}

The quantum theory of a single-mode optical field is based on the Hilbert space   
${\cal H}$ spanned by the photon number states $\{\ket n\,|\,n=0,1,2,...\}$.
We define the usual lowering, raising, and number operators,
$
a:=\sum_{n=0}^\infty\sqrt{n+1}\kb n{n+1},$ 
$
a^*:=\sum_{n=0}^\infty\sqrt{n+1}\kb{n+1}n,
$
$
N:=a^*a=\sum_{n=0}^\infty n\kb n n,
$ 
respectively.

Coherent states $\ket z:=\rme^{-|z|^2/2}\sum_{n=0}^\infty z^n/\sqrt{n!}\ket n$, $z\in\C$, describe the laser light; here $|z|\in[0,\infty)$ is the energy or intensity parameter and $\arg z\in[0,2\pi)$ is the phase parameter.
The number operator shifts the phase, that is, $\rme^{\rmi\theta N}\ket z=\ket{z\rme^{\rmi\theta}}$.
    
A normalized positive operator measure (POM) $E:\,{\cal B}[0,2\pi)\to\cal L(H)$ 
is a phase (parameter) measurement of laser light if
$$
\langle{ze^{-i\theta}}|E(X)|{ze^{-i\theta}}\rangle=\langle{z}|E(X\dot+\,\theta)|{z}\rangle
$$
for all $z\in\C$, $\theta\in[0,2\pi)$, and $X\in{\cal B}[0,2\pi)$.\footnote{$\cal L(H)$ is the set of bounded operators on $\cal H$, ${\cal B}(\Omega)$ is the Borel $\sigma$-algebra of any topological space $\Omega$, and $\dot+$ means the addition modulo $2\pi$.
A mapping $E:\,{\cal B}(\Omega)\to\cal L(H)$ is a POM if and only if $X\mapsto\<\psi|E(X)\psi\>$ is a probability measure for any vector state $\psi\in\cal H$.}
It is easy to show \cite{LaPe02} that a POM $E$ is a phase measurement if and only if it is phase shift covariant, that is, if
$$
\rme^{\rmi\theta N}E(X)\rme^{-\rmi\theta N}=E(X\dot+\,\theta)
$$
holds for all $X$ and $\theta$.
Hence, we say that a POM $E:\,{\cal B}[0,2\pi)\to\cal L(H)$ is a {\it (covariant) phase observable} if 
it is phase shift covariant.

The structure of phase observables is well known, see e.g.\ \cite{Ho83,LaPe99, CaDeLaPe}. Any phase observable $E$ is of the form
$$
E(X)=\sum_{m,n=0}^\infty c_{m,n}\,\frac{1}{2\pi}\int_X \rme^{\rmi(m-n)\theta}\rmd\theta\,\kb m n
$$
where the (unique) {\it phase matrix} $(c_{m,n})_{m,n=0}^\infty$ is positive semidefinite and $c_{m,m}=1$ for all $m$.
As a positive semidefinite matrix, $(c_{m,n})$ has a Kolmogorov decomposition (see, e.g.\ \cite{CaDeLaPe,Ho84,HoPe}), that is, there exists a sequence of unit vectors 
$(\eta_n)_{n=0}^\infty$ of $\cal H$, such that 
$
c_{m,n}=\langle\eta_m|\eta_n\rangle
$
for all $m,\,n$.  The sequence $(\eta_n)$ is not unique but, by defining a new Hilbert space ${\cal H}_{(\eta_n)}$ as the closure of 
$\lin\{\eta_n\,|\,n=0,1,...\}$, one sees that a certain uniqueness can be reached as follows \cite{HoPe}:
if $(\fii_n)$ is another sequence giving $(c_{m,n})$
and  ${\cal H}_{(\fii_n)}$ as above, then there exists a unitary operator $U:\,{\cal H}_{(\eta_n)}\to{\cal H}_{(\fii_n)}$
such that $U\eta_n=\fii_n$ for all $n$.
Especially, the dimension of ${\cal H}_{(\eta_n)}$ depends only on $(c_{m,n})$ and we may define
the {\it rank} of $(c_{m,n})$ (or $E$) as $\dim{\cal H}_{(\eta_n)}$.
We denote it by $\rank E$.

In what follows, we consider always a {\it minimal} Kolmogorov decomposition of a phase matrix $(c_{m,n})$, that is, a unit vector sequence $(\eta_n)$ of a Hilbert space ${\cal K}$ such that $c_{m,n}=\langle\eta_m|\eta_n\rangle$ for all $m,\,n$ and vectors $\eta_n$ span $\cal K$. Then $\rank E=\dim\cal K$.

\subsection{The canonical phase measurement}

The canonical phase observable $E_{\mathrm{can}}$ is determined by the phase matrix with the elements $c_{n,m}\equiv1$ \cite{Ho82,LaPe00}. Its minimal Kolmogorov decomposition is given by a constant vector sequence $\eta_n\equiv\eta\in\cal H$ so that $\cal K=\C\eta\cong\C$. Hence, $\rank E_{\mathrm{can}}=\dim\C\eta=1$. 

The canonical phase observable is associated to the polar decomposition of the lowering operator $a$, that is,  
$$
a=\int_0^{2\pi}\rme^{\rmi\theta}\rmd E_{\mathrm{can}}(\theta)\sqrt{N}.
$$
Moreover, $E_{\mathrm{can}}$ is (up to a unitary equivalence)
the only phase observable which generates number shifts \cite{LaPe00}. This suggests that the number operator $N$ and the canonical phase $E_{\mathrm{can}}$ form a canonical pair as the position and momentum observables.

For any phase observable $E$, let $g^E_z$ be the probability density of the coherent state phase measurement, that is,
$$
\langle z|E(X)|z\rangle\equiv
\frac{1}{2\pi}\int_Xg^E_z(\theta)\rmd\theta.
$$
Now the canonical measurement $E_{\mathrm{can}}$ gives the highest peak: 
$$
g^E_z(\arg z)\le g^{E_{\mathrm{can}}}_z(\arg z).
$$
In addition, $g^{E_{\mathrm{can}}}_z$
tends to the $2\pi$-periodic Dirac $\delta$-distribution in the classical limit $|z|\to\infty$ and for sufficiently large energies $|z|$, we have the approximative uncertainty relation
$$
\Delta_{|z\rangle} E_{\mathrm{can}}\Delta_{|z\rangle}N\approx\frac{1}{2}.
$$ 
where $\Delta_{|z\rangle}$ are the square roots of (minimum) variances \cite{LaMa,LaPe00,Pe}.
All these facts demonstrate the canonicity of $E_{\mathrm{can}}$ (for more properties of $E_{\mathrm{can}}$, see the list in page 51 of \cite{Pe}).
  
\section{Extreme phase measurements}

The set of phase observables is convex meaning that, for any two phase observables $E_1$ and $E_2$, one can form a (random mixture) phase observable $E=\lambda E_1+(1-\lambda)E_2$ where $0\le \lambda\le 1$.
A phase observable $E$ is {\it exteme} or {\it pure} if it does not allow nontrivial convex decompositions, that is, if
$E=\lambda E_1+(1-\lambda)E_2$ 
implies that $E_1=E_2=E$. If $E$ is extreme then the coherent state phase statistics $g_z^E$  cannot be obtained by measuring other phase observables in the coherent state $\ket z$ and then mixing their statistics.
The following theorem \cite{HoPe,KiPe} characterizes extreme phase observables.  

Let $E$ be a phase observable associated to unit vectors $\eta_n\in\cal K$ which span $\cal K$.
  
\begin{theorem}
$E$ is extreme if and only if, for any bounded operator $A:\,\cal K\to\cal K$,
$$
\langle\eta_n|A\eta_n\rangle=0\hspace{1cm}{\rm for\;all\;}n\in\{0,1,2,...\},
$$
implies that $A=0$.
\end{theorem}

The next theorem \cite{KiPe} shows that there exist infinite number of extreme phase observables.

\begin{theorem}
There exist extreme phase observables of any rank $\in\{1,2,...,\infty\}$.
\end{theorem}

Since the canonical phase $E_{\rm can}$ is of rank 1, it is automatically extreme \cite{HoPe}.\footnote{In the case of the canonical phase, $\eta_n\equiv\eta$ and $\cal K=\C\eta$.
For any $A=a\kb\eta\eta$ the condition $\langle\eta|A\eta\rangle=a=0$ implies that $A=0$.}
Other rank 1 phase observables are unitarily equivalent to $E_{\rm can}$, that is, they are of the form $U^*E_{\rm can}U$ 
where the unitary operator $U$ commutes with the representation $\theta\mapsto\rme^{\rmi\theta N}$ of U(1), that is, $U$ is diagonal in the number basis. Indeed, if $E$ is of rank 1, the Hilbert space $\cal K$ (associated to the minimal Kolmogorov decomposition) can be chosen to be $\C$. Thus, the unit vector sequence $(\eta_n)$ is just a sequence of complex numbers $\rme^{\rmi\alpha_n}$, $\alpha_n\in[0,2\pi)$, and $U=\sum_{n=0}^\infty\rme^{\rmi\alpha_n}|n\rangle\langle n|$.

Recently, we have proved \cite{HePe} the following stronger result:

\begin{theorem}
The canonical phase $E_{\rm can}$ is extreme in the convex set of \emph{all} POMs ${\cal B}(\R)\to\cal L(H)$.
\end{theorem}

This condition supports the canonicity of $E_{\rm can}$; this result has been known to be true for spectral measures.\footnote{For spectral measures this is obvious since projections are extremals in the convex set of effects.}
  
\begin{remark}\rm
There is no realistic direct measurement scheme for $E_{\rm can}$ but some other phase observables, so-called phase space phase observables \cite{LaPe99}, can be measured. Let $D(z):=\rme^{za^*-\overline{z}a}$,  $z\in\mathbb C$, be the displacement operator and $T:=\sum_{n=0}^\infty\lambda_n|n\rangle\langle n|$ where $\lambda_n\ge 0$ for all $n$ and $\sum_{n=0}^\infty\lambda_n=1$. A phase space phase observable $E_T$ is defined by
$$
E_T(X):=\frac{1}{\pi}\int_X\int_0^\infty D(r\rme^{\rmi\theta})T D(r\rme^{\rmi\theta})^*r\,\rmd r\,\rmd \theta.
$$
In principle, any phase space phase observable can be measured by using an eight-port homodyne detector \cite{Le,KiLa}.
Indeed, $E_{\kb00}$ has been measured by Walker and Carroll \cite{WaCa}.
It can be shown \cite{CaHePeTo09} that the rank of any $E_T$ is $\infty$ and $E_T$ is not extreme. This suggests that better phase measurement schemes could be found in future.
\end{remark}

\section{Extreme rotated quadratures}
Define the quadrature operators $Q:=(a^*+a)/\sqrt{2}$ and
$P:=(a^*-a)i/\sqrt{2}$ which, in the coordinate
representation ${\cal H}\cong L^2(\R)$, are the usual position and momentum operators
$(Q\psi)(x) = x\psi(x)$ and $(P\psi)(x)= -i{\rmd\psi(x)}/{\rmd x}$, respectively (in units where $\hbar=1$).

For any $\theta\in [0,2\pi)$, define the
rotated quadrature operators $Q_\theta$ and $P_\theta$ by
\begin{eqnarray*}
Q_\theta &:=& R(\theta)QR(\theta)^*, \\
P_\theta &:=& R(\theta)PR(\theta)^*
\end{eqnarray*} 
where $R(\theta):=\rme^{\rmi\theta N}$.
Note that $P_\theta=Q_{\theta+\pi/2}$ and $R(\pi/2)$ is the Fourier-Plancherel operator.
The rotated quadratures can be measured by balanced homodyne detection \cite{Le,KiLa08}.
Next we define fuzzy rotated quadratures as the solutions of a covariance system.

Fix $\theta\in [0,2\pi)$ and choose a rotated momentum representation of ${\cal H}\cong L^2(\R)$ such that\begin{eqnarray*}
(Q_\theta\varphi)(p) &=& i{\rmd\varphi(p)}/{\rmd p}, \\ 
(P_\theta\varphi)(p) &=& p\varphi(p). 
\end{eqnarray*} 
 A POM $F_\theta:\,{\cal B}(\R)\to{\cal L(H)}$ is a {\it fuzzy rotated quadrature} if
\begin{equation}\label{cov}
\rme^{\rmi q P_\theta}F_\theta(X)e^{-\rmi q P_\theta}=F_\theta(X+q)
\end{equation}
for all $q\in\R$ and $X\in{\cal B}(\R)$.
Any $F_\theta$ is of the form 
$$
\langle\varphi|F_\theta(X)\psi\rangle=
\frac{1}{2\pi}\int_X\int_\R\int_\R \rme^{\rmi(p-p')x}\langle\eta_p|\eta_{p'}\rangle\overline{\varphi(p)}\psi(p')\rmd p\,\rmd p'\rmd x
$$
for all integrable $\varphi,\,\psi\in L^2(\R)$,
where $p\mapsto\eta_p\in\cal H$ is a (non-unique measurable) family of unit vectors \cite{Ho83,Ho84,HoPe}.
Let $\cal K$ be the closure of the image of the mapping
$$
L^1(\R)\cap L^2(\R)\ni \varphi\mapsto\int_\R \varphi(p)\eta_p\rmd p\in\cal H.
$$
One may assume that $\eta_p\in\cal K$ for all $p\in\R$ \cite{HoPe}.\footnote{The choice $\cal K$ gives a minimal Kolmogorov decomposition for a certain positive measurable field of operators. It is unique up to a unitary transformation \cite{HoPe}.}
Define the (unique) rank of $F_\theta$ as 
$$
\rank F_\theta:=\dim\cal K.
$$

It is easy to see that, for a fixed $\theta$, fuzzy rotated quadratures form a convex set.
Similarly as in the case of phase, we have the following theorems \cite{HoPe}:
\begin{theorem}
$F_\theta$ is extreme if and only if, for any bounded operator $A:\,\cal K\to\cal K$,
$$
\langle\eta_p|A\eta_p\rangle=0\hspace{1cm}{\rm for\;almost\;all\;}p\in\R,
$$
implies that $A=0$.
\end{theorem}

\begin{theorem}
There exist extreme fuzzy rotated quadratures of any rank $\in\{1,2,...,\infty\}$.
\end{theorem}
Moreover, any $F_\theta$ of the rank 1 is a spectral measure, extreme, and unitarily equivalent to the sharp quadrature observable $Q_\theta$ (for which $\langle\eta_p|\eta_{p'}\rangle\equiv1$), that is, 
$$
\langle\varphi|F_\theta(X)\psi\rangle=
\frac{1}{2\pi}\int_X\int_\R\int_\R \rme^{\rmi(p-p')x}\rme^{\rmi(\alpha_p-\alpha_{p'})}\overline{\varphi(p)}\psi(p')\rmd p\,\rmd p'\rmd x
$$
where $\alpha_p\in[0,2\pi)$ and the unitary operator $U$ is given by $(U\psi)(p):=\rme^{\rmi\alpha_p}\psi(p)$.
As a spectral measure, $Q_\theta$ is extremal in the convex set of all POMs $\cal B(\R)\to\cal L(H)$.
If the rank $F_\theta>1$ then $F_\theta$ cannot be a spectral measure \cite{HoPe}.

\begin{remark}\rm
If, in addition to the covariance condition (\ref{cov}), a fuzzy rotated quadarature $F_\theta$ satisfies the invariance condition
$$
\rme^{\rmi p Q_\theta}F_\theta(X)e^{-\rmi p Q_\theta}=F_\theta(X)
$$
for all $p$ and $X$, then it is the following convolution:
$$
F_\theta(X)=\int_\R\rho(X-x)\rmd \Pi_{Q_\theta}(x)
$$
where $\rho$ is a probability measure on $\R$ and $\Pi_{Q_\theta}$ is a spectral measure of $Q_\theta$ \cite{CaHeTo}.
Hence, $F_\theta$ is then a postprocessing of $Q_\theta$ \cite{CaHePeTo09}.
\end{remark}

\section{Discussion}
Since there is no phase shift covariant spectral measures (self-adjoint operators) \cite{LaPe99}, 
the quantum phase problem is a true example of the case where the conventional formulation of quantum mechanics,
where observables are self-adjoint operators, cannot be sufficient. 
We have seen that some properties of $Q_\theta$ and $E_{\rm can}$ correspond each other except that $E_{\rm can}$ is not a spectral measure and thus a conventional observable.
It should be stressed that, since $E_{\rm can}$ is extreme in the set of all observables, it cannot be considered as a noisy measurement of {\it any} spectral measure \cite{HePe}. This underlines the canonicity of $E_{\rm can}$.

The results of this paper can be generalized for (almost) any observables, that is, for  POMs. 
Various classes of observables correspond to the solutions of covariance systems with particular symmetry groups associated to them \cite{Ho82,Ho87}. But quite rarely covariance systems admit spectral measure solutions.
As Holevo suggests in \cite{Ho87}, the canonical quantization must be generalized to the context of covariance systems. In the same paper, he solves covariance systems in the case of type I symmetry groups. The most used symmetry groups in physics are of type I, so that the characterization is quite extensive.

As we have seen, a covariance system may have infinite number of solutions (covariant observables), so that it is important to find the physically most reasonable ones. Since covariant POMs form a convex set, its extremals are good candidates for these observables (they describe pure measurements). In \cite{CaHePeTo08,ChDA,DA,HoPe}, extremals are characterized for rather broad classes of covariance systems. 

The final problem is to find 'canonical' observable(s) from the set of extremals. As shown in this paper, it is possible for phase observables although there is no projection valued phase observables at all. 

\ack 
The author thanks Pekka Lahti for the carefully reading of the manuscript.

\end{document}